\documentclass[osajnl,twocolumn,showpacs,superscriptaddress,10pt]{revtex4-1} 

\pdfoutput=1

\usepackage{amsmath,amssymb,graphicx}
\usepackage[colorlinks=true, allcolors=blue]{hyperref}

\begin{document}

\title{Continuous-wave coherent Raman spectroscopy for improving the accuracy of Raman shifts}

\author{Hugo Kerdoncuff}\email{Corresponding author: hk@dfm.dk}
\affiliation{Danish Fundamental Metrology, Kogle Alle 5, DK-2970 H\o rsholm, Denmark}

\author{Mikael Lassen}
\affiliation{Danish Fundamental Metrology, Kogle Alle 5, DK-2970 H\o rsholm, Denmark}

\author{Jan C. Petersen}
\affiliation{Danish Fundamental Metrology, Kogle Alle 5, DK-2970 H\o rsholm, Denmark}

%\dates{Compiled \today}

%\ociscodes{(300.6450) Raman spectroscopy, (290.5910) Stimulated Raman scattering, (230.5440) Polarization-selective devices}

\begin{abstract}
Raman spectroscopy is an appealing technique that probes molecular vibrations in a wide variety of materials with virtually no sample preparation. However, accurate and reliable Raman measurements are still a challenge and require more robust and practical calibration methods. We demonstrate the implementation of a simple low-cost continuous-wave stimulated Raman spectroscopy scheme for accurate and high-resolution spectroscopy.
We perform shot noise limited continuous-wave stimulated Raman scattering (cwSRS) as well as continuous-wave coherent anti-Stokes Raman scattering (cwCARS) on polystyrene samples.
Our method enables accurate determination of Raman shifts with an uncertainty below 0.1 cm\textsuperscript{-1}. The setup is used for the characterization of reference materials required for the calibration of Raman spectrometers. Compared with existing standards, we provide an order of magnitude improvement of the uncertainty of Raman energy shifts in a polystyrene reference material.

\end{abstract}

%\begin{document}

\maketitle

There is an increasing demand for novel spectroscopic techniques for improved quality assurance and validation of products from pharmaceutical, cosmetic, food, and agricultural industries \cite{Das2011, Yang2011, Edinger2018, Duraipandian2019}. Raman spectroscopy is a powerful technique since it is highly versatile and non-invasive. It typically determines vibrational modes of molecules, although rotational and other low-frequency modes may also be observed. Infrared absorption spectroscopy, such as Fourier-transform infrared spectroscopy (FTIR), is routinely used for quantitative analysis of material while Raman spectroscopy mainly has been used as a qualitative measurement method. Only in recent years has there been an increased effort to develop the technique into a quantitative method \cite{Pelletier2003, Johansson2007, Strachan2007, Bell2008, Duraipandian2018}. In order to produce accurate, reliable and repeatable results, Raman spectrometers require frequent calibration that is time consuming and very cumbersome in routine operation. The calibration of the spectrometer wavelength axis commonly uses emission lines from gas-discharge lamps (e.g. mercury, neon, argon) whose wavelengths are well-defined with very low uncertainties. However, placing the lamp at the sample position is often impractical and calibrating the Raman shift wavenumber further requires knowledge of the excitation wavelength. Therefore, it has become customary to use reference samples with predefined Raman shifts regardless of the excitation wavelength \cite{ASTM, Hutsebaut2005, Rodriguez2011, Bocklitz2015}. There exists a variety of reference materials suitable for various applications and wavenumber range, such as silicon, paracetamol, toluene, cyclohexane and polystyrene \cite{ASTM}. Among those, polymers are particularly interesting due to their wide availability, low cost, ease of handling and chemical stability \cite{Liu2018}. To be applicable for calibration procedures, reference materials must be carefully characterized and provide high spectral accuracy. This is typically accomplished using high-resolution Fourier-transform Raman spectrometers \cite{ASTM}. However, the technique measures interferograms that are not directly readable and it requires additional data processing to obtain Raman spectra. We present an alternative and novel method for the acquisition of high-resolution and accurate Raman spectra of polystyrene reference materials using shot noise limited continuous-wave stimulated Raman scattering (cwSRS).

Stimulated Raman scattering (SRS) refers to the coherent scattering of photons from a high-frequency field (pump) to a low-frequency field (probe) that happens when the frequency detuning between the two fields matches a Raman transition. Its counterpart technique, coherent anti-Stokes Raman scattering (CARS), refers to the coherent scattering to higher frequency than the pump field under similar matching conditions of pump and probe. Most realization of SRS and CARS uses two synchronized picosecond or femtosecond pulsed lasers that deliver very high peak powers to drive the Raman transition \cite{Cheng2013}. While it strongly enhances the Raman scattering, it also increases the possibility of radiation damage from nonlinear effects such as two-photon absorption \cite{Konig1995, Konig1999, Fu2006}. Moreover, the short pulse duration limits the spectral resolution to a few cm\textsuperscript{-1}. On the other hand, cwSRS reduces sample damage due to the lower peak intensity, achieves high resolution due to the narrow linewidth of continuous-wave (cw) lasers \cite{Owyoung1977, Owyoung1978, Domenech2013, Westergaard2015}, and decreases cost and complexity of the setup \cite{Meng2013, Hu2013}. In this study, we benefit from all the advantages of cwSRS spectroscopy to implement high-resolution and accurate measurements of Raman spectra of polymer materials. Our system operates within the fingerprint region of organic molecules (500 -- 1800 cm\textsuperscript{-1}), which is particularly relevant for biological and medical applications. We achieve one order of magnitude improvement on the accuracy of the measured Raman shifts compared to the standard of the American Society for Testing and Materials (ASTM) \cite{ASTM}.

\begin{figure}[htbp]
\centering
\includegraphics[width=\linewidth]{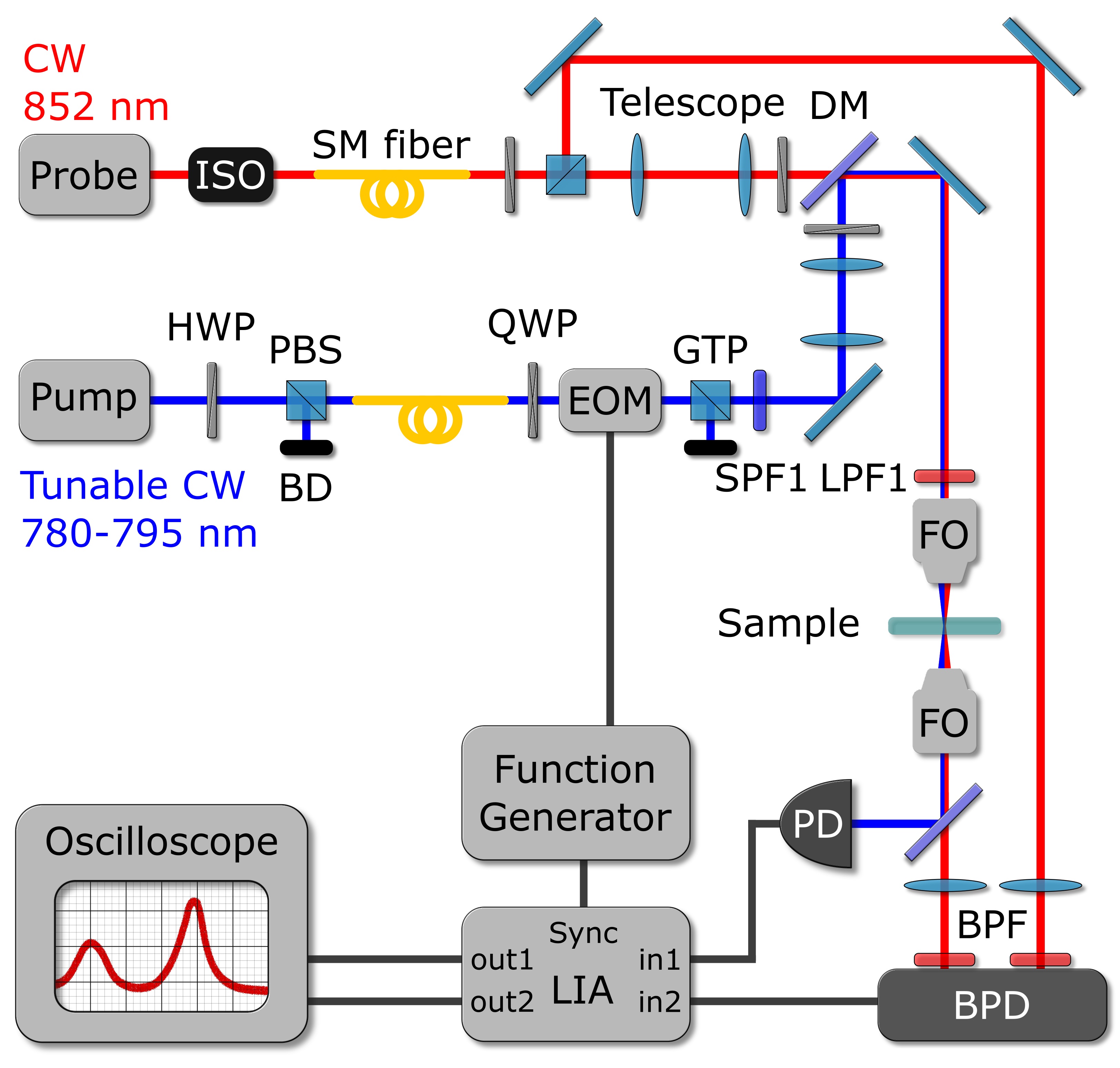}
\caption{cwSRS setup configured to measure SRG on the probe beam. The red and blue lines represent the probe and pump beams, respectively. ISO, isolator; SM, single mode; DM, dichroic mirror; LPF, longpass filter; SPF,shortpass filter; BPF, bandpass filter; HWP, half-wave plate; QWP, quarter-wave plate; PBS, polarizing beam splitter; GTP, Glan-Taylor prism; EOM, electro-optic modulator; BD, beam dump; FO, focusing objective; (B)PD, (Balanced) photodetector; LIA, Lock-in amplifier.}
\label{fig:setup}
\end{figure}

Figure \ref{fig:setup} shows our cwSRS setup configured to measure stimulated Raman gain (SRG) on the probe beam. The pump beam is provided by a tunable external cavity diode laser with a tapered amplifier (Toptica TA pro) and the probe beam by a single-frequency grating stabilized laser diode (LD852-SEV600). The combination of the fixed probe wavelength at 852 nm and the tunable pump wavelength covering the range 781 nm to 787 nm allows SRS operation from 968 cm\textsuperscript{-1} to 1060 cm\textsuperscript{-1}. The probe diode laser is protected from back reflections by an optical isolator in order to provide optical power stability. The spatial modes of the pump and probe beams are cleaned with single mode fibers. The two beams are mode matched by two telescopes before being combined on a longpass dichroic mirror (Semrock LPD02-785RU). The polarization of the two beams are controlled by polarizing beam splitters (PBS) and wave plates. A 20$\times$ microscope objective with 0.20 NA focuses the beams on the sample. Prior to positioning the sample at the focal point, the overlap of the probe and pump beams is optimized by placing a plane-cut single mode fiber at the sample position and optimizing the coupling of the two beams into the fiber. This simple method readily provides optimum alignment and mode matching conditions for SRS after replacing the fiber tip by the sample. The probe and pump beams are collected after the sample by a 20$\times$ microscope objective with 0.20 NA. The objective is mounted on a 3-axis translation stage in order to optimize collection of probe light that carries the SRG signal. After collection, the pump and probe beams are separated by a longpass dichroic mirror. The reflected pump beam is directed to a fast PIN photodetector and the probe beam to a homebuilt balanced photodetector.

\begin{figure}[tbp]
\centering
\includegraphics[width=\linewidth]{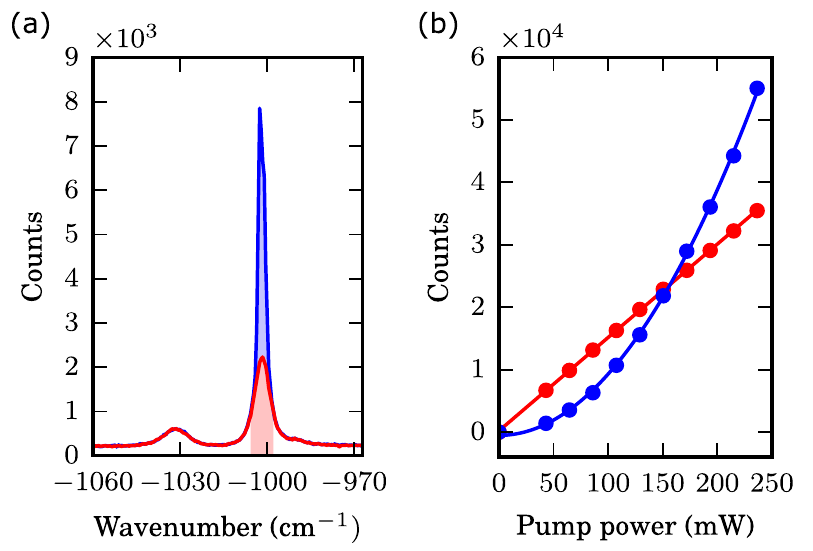}
\caption{{\bf (a)} Spectrum of anti-Stokes scattering from a polystyrene sample without (red trace) and with cwCARS excitation of the breathing mode of the phenyl ring (blue trace). Here pump and probe power at the sample are 172 mW and 65 mW, respectively. Summing the counts over the shaded areas for different pump power shows the scaling of ARS (red dots) and CARS (blue dots) in {\bf (b)}. Fits to the data demonstrate linear (red trace) and quadratic (blue trace) scaling of the ARS and CARS with pump power, respectively. Measurement uncertainties are within the scatter plot symbols.}
\label{fig:cars}
\end{figure}

In the absence of the probe beam, spontaneous Raman scattering (SpRS) can be observed by directing the scattered Stokes light to a Raman spectrometer (iHR320, Horiba) via a multimode fiber. In order to check the tuning of the probe wavelength relative to the Raman transitions we observed the spontaneous Stokes scattered light and the probe beam simultaneously with the spectrometer. In addition, we detect the anti-Stokes Raman scattering (ARS) and CARS by substituting the longpass dichroic mirror by a shortpass filter. CARS was observed as an enhancement of ARS at the Raman shift matching the pump-probe detuning. The CARS signal was used for checking and optimizing the alignment and mode matching of the pump and probe beams onto the sample. In Figure \ref{fig:cars} (a), the pump and probe wavelengths are tuned to the breathing mode of the phenyl ring in polystyrene. With pump and probe powers of 172 mW and 65 mW, respectively, the collected CARS and ARS have similar intensities. We verify the square power law of CARS as a function of pump power (see Figure \ref{fig:cars} (b)), which explains the tremendous signal enhancement from CARS over ARS when using the high peak powers of pico- and femtosecond lasers \cite{Cheng2013}.  Even though cwCARS has been applied in the past for high-resolution spectroscopy of gases and liquids \cite{Barrett1975, Akhmanov1978}, we report here the first observation of cwCARS in a solid sample to our knowledge.

In order to observe SRG, the pump intensity is modulated before combination with the probe beam by using a resonant electro-optic amplitude modulator (Thorlabs EO-AM-R-10.7-C1) and a Glan-Taylor polarizer. The modulator is driven at 10.430 MHz with a sinusoidal modulation, which generates a SRG signal at the same probe laser sideband frequency of 10.43 MHz. We apply balanced detection to measure SRG with a sensitivity at the shot noise limit. A reference beam is tapped off from the probe beam before reaching the sample and its noise is subtracted from the probe beam that carries the SRG signal after the sample. Appropriate balancing of the intensities of the reference and signal beams enables reduction of classical noise from the diode laser, with a common mode rejection of more than 20 dB of the photocurrent at 10.43 MHz. The SRG signal and noise are then measured with an electronical spectrum analyzer (Agilent E4411B).

\begin{figure}[htbp]
\centering
\includegraphics[width=\linewidth]{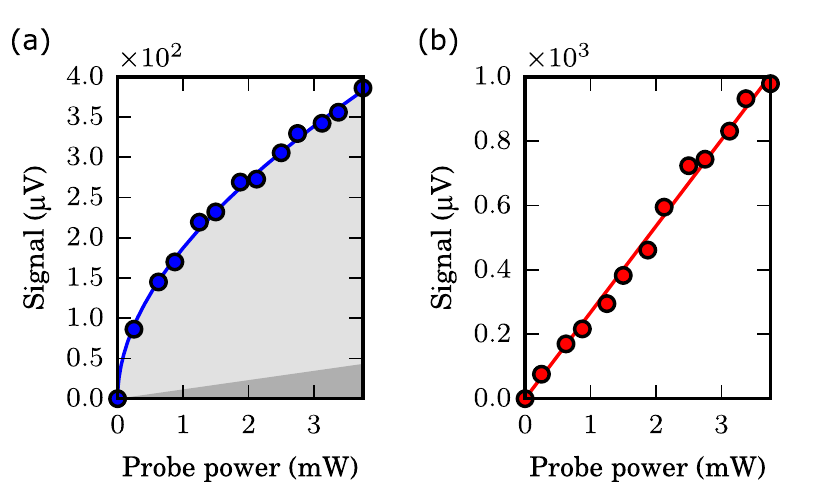}
\caption{Scaling of the measurement noise {\bf (a)} and SRS signal {\bf (b)} with optical probe power. The measured noise (blue dots) is fitted with the curve $y = 11.3 x + 175.9 \sqrt{x}$ (blue trace). The dark grey area indicates the linear contribution from classical noise whereas the light grey area indicates quantum shot noise that scales with the square root of the probe power. The measured SRS signal amplitude (red dots) is fitted with a straight line (red trace). Measurement uncertainties are within the scatter plot symbols.}
\label{fig:SRS_scaling}
\end{figure}

Figure \ref{fig:SRS_scaling} (a) shows the scaling of the measurement noise with optical probe power. Measurements were performed with 106 mW of average pump power at the sample, while varying the probe power from 0 to 3.8 mW. The signal and noise are measured using a spectrum analyzer with a resolution and video bandwidth of 1 kHz and 30 Hz, respectively. The noise is dominated by optical shot noise ($>$ 87 \%) in the range of probe power that are used ($<$ 3.8 mW). A fit to the measured noise floor gives a scaling of $y = 11.3 x + 175.9 \sqrt{x}$, with linear and square root scaling indicating classical noise and quantum shot noise, respectively. The classical noise is due to electrical pickup noise \cite{Hu2013}. The expected linear scaling of the SRS signal with probe power is shown in Figure \ref{fig:SRS_scaling} (b). Measurement at  the shot noise limit is rendered possible by balanced detection at the sideband frequency away from the main technical noise, despite the use of a cheap laser diode as probe source.

\begin{figure}[htbp]
\centering
\includegraphics[width=\linewidth]{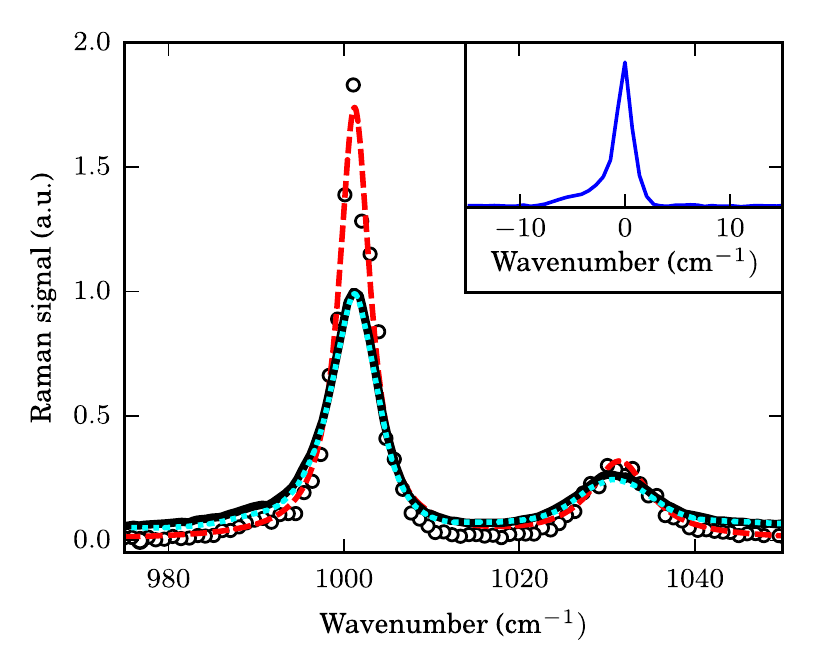}
\caption{Comparison of SRS and SpRS spectra. The measured SRS spectrum (black circles) is fitted with a sum of two lorentzians (red dashed line). The convolution (cyan dotted line) of the SRS spectrum with the slit function of the Raman spectrometer (inset) is compared to a SpRS spectrum (black line). The spectra are normalized to their respective area under the curve then scaled relative to the peak maximum of the SpRS spectrum.}
\label{fig:srs_sprs}
\end{figure}

SRS spectra were obtained by scanning the pump laser wavelength. The spectra in Figure \ref{fig:srs_sprs} were measured on a 1.5 mm-thick polystyrene slab in the range 968 cm\textsuperscript{-1} to 1060 cm\textsuperscript{-1}. The black line corresponds to a SpRS spectrum and the black circles denote the measured SRS spectrum. The latter is fitted with a sum of two Lorentzians, resulting in the dashed red line. The cyan dotted line is the convolution of the SRS spectrum with the slit function of the Raman spectrometer for SpRS, shown in inset. It matches the SpRS spectrum after normalizing with the area under the curve. This denotes the high-resolution capability of cwSRS compared to grating-based SpRS instruments. cwSRS is therefore capable of providing more accurate values for peak heights and linewidths of Raman transitions. With cwSRS the spectral resolution is only limited by the linewidths of the probe and pump lasers. In our experiment, the linewidths of the probe and pump lasers are 20 MHz and 100 kHz respectively, giving a resolution of $7\cdot 10^{-2}$ cm\textsuperscript{-1}.

\begin{figure}[htbp]
\centering
\includegraphics[width=\linewidth]{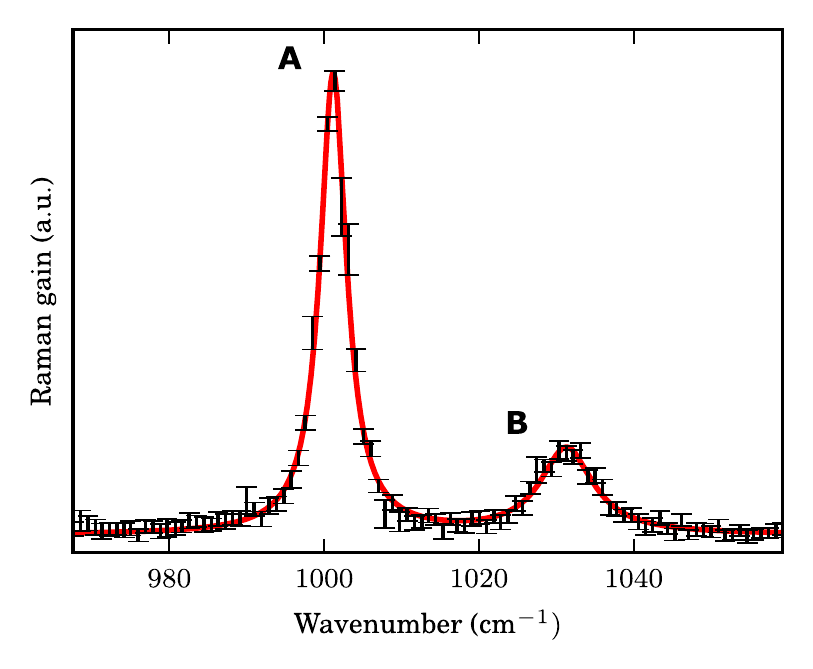}
\caption{SRS spectrum of a NIST reference material for FTIR made of polystyrene. Error bars indicate standard uncertainties and the red trace is a fit to the measured data with a sum of two Lorentzians. The fitting is done by the least squares method \cite{Nielsen2002} and the results are shown in Table \ref{tab:results}.}
\label{fig:srs_gum}
\end{figure}

As metrology approach and demonstration of Raman spectroscopy we apply our SRS system to high-accuracy determination of Raman shifts in a reference material. The advantage of the SRS method for this purpose is the ability to accurately infer the Raman shift via measurement of the probe and pump laser wavelengths by a wavemeter with sub-picometer precision. We measure the laser wavelengths with a wavemeter (HP 86120B) calibrated to a frequency standard with an uncertainty of 0.6 pm. We report values of wavelengths and wavenumbers in vacuum, unless stated otherwise. The Raman spectrum is obtained by scanning the pump wavelength by steps of 60 pm, equivalent to about 0.9 cm\textsuperscript{-1}. At each step, the SRG signal is acquired on a digital spectrum analyzer (Moku:Lab) and normalized to the modulation amplitude of the pump intensity. The resulting values plotted against the Raman shift calculated from the wavelength measurements give the Raman spectrum of the sample. Figure \ref{fig:srs_gum} shows the measured Raman spectrum of a NIST traceable reference material for FTIR made of polystyrene (NIST\textsuperscript{\textregistered} SRM\textsuperscript{\textregistered} 1921b). The values of the Raman gain is averaged over a hundred measurements and the uncertainties (error bars) are set as a minimum of twice the standard deviation of the means. The data are fitted with a sum of two Lorentzians using the least squares method \cite{Nielsen2002}. The standard uncertainty of a few measurement values are adjusted in order to obtain a reasonable goodness of fit according to the $\chi^2$ test. We calculate $\chi_{obs}^2 = 70.4$ for a number of degrees of freedom $\delta = 92$, giving a probability $P\{\chi^2(\delta)>\chi_{obs}^2\} = 95 \%$ that indicates good matching between our measurement model and our data within the measurement uncertainties. The results of our model fit are listed in Table \ref{tab:results}. The wavenumber values in air are calculated from the wavelength values in air that are converted using the Ciddor equation \cite{Ciddor1996} for the refractive index in dry air at 15 $^{\circ}$C, 101.325 kPa and with 450 ppm CO\textsubscript{2} content. Compared with the existing standard by ASTM International \cite{ASTM}, we achieve up to one order of magnitude improvement of the uncertainty of the measured Raman shifts. Furthermore, we measure a peak intensity ratio $I_B/I_A = 0.186 \pm 0.009$ in contrast to the uncorrected value of $0.27$ given by the standard. SRS is directly proportional to the imaginary part of the $\chi^{(3)}$ nonlinear susceptibility and depends only on the difference frequency of the pump and Stokes beams unlike SpRS \cite{Cheng2013}. Therefore, SRS allows for an arbitrary choice of the pump and probe wavelengths to characterize a material.

\begin{table}[htbp]
\centering
\caption{\bf Raman peaks of polystyrene}
\begin{tabular}{ccccc}
\hline
Ref. & $\nu_A$ [cm\textsuperscript{-1}] & $u(\nu_A)$ & $\nu_B$ [cm\textsuperscript{-1}] & $u(\nu_B)$ \\
\hline
vacuum & 1001.19 & 0.04 & 1031.33 & 0.19 \\
air & 1001.46 & 0.04 & 1031.61 & 0.19 \\
ASTM \cite{ASTM} & 1001.4 & 0.54 & 1031.8 & 0.43 \\
\hline
\end{tabular}
\label{tab:results}
\end{table}

In summary, we demonstrated cwSRS with a low-cost setup performing shot noise limited spectral measurements. We also report for the first time to our knowledge observation of cwCARS on a solid material. cwSRS has several advantages over current pulsed SRS schemes. The very high monochromaticity of cw lasers provides high spectral resolution as well as accurate frequency references from which Raman shifts can be inferred with high accuracy. We applied high-resolution cwSRS to measure the Raman spectrum of a polystyrene reference material and followed the GUM \cite{GUM} method to calculate values and uncertainties of the Raman shifts of two molecular vibrational modes. We achieve up to one order of magnitude improvement of the accuracy of the measured Raman shifts compared to the ASTM standard. Accurate determination of Raman shifts is particularly important for precise calibration of Raman shift axis based on standard scatterers \cite{Hutsebaut2005, Rodriguez2011, Bocklitz2015}. The standard scatterer calibration method is more practical and rapid and does not require knowledge of the frequency of the excitation laser. Therefore, our method contributes to the wider development and application of Raman spectroscopy in industrial settings, where effectiveness and efficiency are of primary importance.

\section*{Acknowledgment}
%\section*{Acknowledgment}

We acknowledge the financial support from the Danish Agency for Institutions and Educational Grants.

The authors thank L. Nielsen for assistance with the uncertainty analysis, and D. Tsikritsis for interesting discussion about the experimental setup.

%\section{References}

%%%%%%%%%%%%%%%%%%%%%%% References %%%%%%%%%%%%%%%%%%%%%%%%%

% Bibliography
\bibliography{20190919_cwSRS_AM_arXiv_biblio}

\begin{thebibliography}{29}%
\makeatletter
\providecommand \@ifxundefined [1]{%
 \@ifx{#1\undefined}
}%
\providecommand \@ifnum [1]{%
 \ifnum #1\expandafter \@firstoftwo
 \else \expandafter \@secondoftwo
 \fi
}%
\providecommand \@ifx [1]{%
 \ifx #1\expandafter \@firstoftwo
 \else \expandafter \@secondoftwo
 \fi
}%
\providecommand \natexlab [1]{#1}%
\providecommand \enquote  [1]{``#1''}%
\providecommand \bibnamefont  [1]{#1}%
\providecommand \bibfnamefont [1]{#1}%
\providecommand \citenamefont [1]{#1}%
\providecommand \href@noop [0]{\@secondoftwo}%
\providecommand \href [0]{\begingroup \@sanitize@url \@href}%
\providecommand \@href[1]{\@@startlink{#1}\@@href}%
\providecommand \@@href[1]{\endgroup#1\@@endlink}%
\providecommand \@sanitize@url [0]{\catcode `\\12\catcode `\$12\catcode
  `\&12\catcode `\#12\catcode `\^12\catcode `\_12\catcode `\%12\relax}%
\providecommand \@@startlink[1]{}%
\providecommand \@@endlink[0]{}%
\providecommand \url  [0]{\begingroup\@sanitize@url \@url }%
\providecommand \@url [1]{\endgroup\@href {#1}{\urlprefix }}%
\providecommand \urlprefix  [0]{URL }%
\providecommand \Eprint [0]{\href }%
\providecommand \doibase [0]{http://dx.doi.org/}%
\providecommand \selectlanguage [0]{\@gobble}%
\providecommand \bibinfo  [0]{\@secondoftwo}%
\providecommand \bibfield  [0]{\@secondoftwo}%
\providecommand \translation [1]{[#1]}%
\providecommand \BibitemOpen [0]{}%
\providecommand \bibitemStop [0]{}%
\providecommand \bibitemNoStop [0]{.\EOS\space}%
\providecommand \EOS [0]{\spacefactor3000\relax}%
\providecommand \BibitemShut  [1]{\csname bibitem#1\endcsname}%
\let\auto@bib@innerbib\@empty
%</preamble>
\bibitem [{\citenamefont {Das}\ and\ \citenamefont {Agrawal}(2011)}]{Das2011}%
  \BibitemOpen
  \bibfield  {author} {\bibinfo {author} {\bibfnamefont {R.~S.}\ \bibnamefont
  {Das}}\ and\ \bibinfo {author} {\bibfnamefont {Y.~K.}\ \bibnamefont
  {Agrawal}},\ }\href {\doibase 10.1016/j.vibspec.2011.08.003} {\bibfield
  {journal} {\bibinfo  {journal} {Vibrational Spectroscopy}\ }\textbf {\bibinfo
  {volume} {57}},\ \bibinfo {pages} {163} (\bibinfo {year} {2011})}\BibitemShut
  {NoStop}%
\bibitem [{\citenamefont {Yang}\ and\ \citenamefont {Ying}(2011)}]{Yang2011}%
  \BibitemOpen
  \bibfield  {author} {\bibinfo {author} {\bibfnamefont {D.}~\bibnamefont
  {Yang}}\ and\ \bibinfo {author} {\bibfnamefont {Y.}~\bibnamefont {Ying}},\
  }\href {\doibase 10.1080/05704928.2011.593216} {\bibfield  {journal}
  {\bibinfo  {journal} {Applied Spectroscopy Reviews}\ }\textbf {\bibinfo
  {volume} {46}},\ \bibinfo {pages} {539} (\bibinfo {year} {2011})}\BibitemShut
  {NoStop}%
\bibitem [{\citenamefont {Edinger}\ \emph {et~al.}(2018)\citenamefont
  {Edinger}, \citenamefont {Knopp}, \citenamefont {Kerdoncuff}, \citenamefont
  {Rantanen}, \citenamefont {Rades},\ and\ \citenamefont
  {L\"{o}bmann}}]{Edinger2018}%
  \BibitemOpen
  \bibfield  {author} {\bibinfo {author} {\bibfnamefont {M.}~\bibnamefont
  {Edinger}}, \bibinfo {author} {\bibfnamefont {M.~M.}\ \bibnamefont {Knopp}},
  \bibinfo {author} {\bibfnamefont {H.}~\bibnamefont {Kerdoncuff}}, \bibinfo
  {author} {\bibfnamefont {J.}~\bibnamefont {Rantanen}}, \bibinfo {author}
  {\bibfnamefont {T.}~\bibnamefont {Rades}}, \ and\ \bibinfo {author}
  {\bibfnamefont {K.}~\bibnamefont {L\"{o}bmann}},\ }\href {\doibase
  10.1016/j.ejps.2018.02.012} {\bibfield  {journal} {\bibinfo  {journal}
  {European Journal of Pharmaceutical Sciences}\ }\textbf {\bibinfo {volume}
  {117}},\ \bibinfo {pages} {62} (\bibinfo {year} {2018})}\BibitemShut
  {NoStop}%
\bibitem [{\citenamefont {Duraipandian}\ \emph {et~al.}(2019)\citenamefont
  {Duraipandian}, \citenamefont {Petersen},\ and\ \citenamefont
  {Lassen}}]{Duraipandian2019}%
  \BibitemOpen
  \bibfield  {author} {\bibinfo {author} {\bibfnamefont {S.}~\bibnamefont
  {Duraipandian}}, \bibinfo {author} {\bibfnamefont {J.~C.}\ \bibnamefont
  {Petersen}}, \ and\ \bibinfo {author} {\bibfnamefont {M.}~\bibnamefont
  {Lassen}},\ }\href {\doibase 10.3390/app9122433} {\bibfield  {journal}
  {\bibinfo  {journal} {Applied Sciences}\ }\textbf {\bibinfo {volume} {9}},\
  \bibinfo {pages} {2433} (\bibinfo {year} {2019})}\BibitemShut {NoStop}%
\bibitem [{\citenamefont {Pelletier}(2003)}]{Pelletier2003}%
  \BibitemOpen
  \bibfield  {author} {\bibinfo {author} {\bibfnamefont {M.~J.}\ \bibnamefont
  {Pelletier}},\ }\href {\doibase 10.1366/000370203321165133} {\bibfield
  {journal} {\bibinfo  {journal} {Applied Spectroscopy}\ }\textbf {\bibinfo
  {volume} {57}},\ \bibinfo {pages} {20A} (\bibinfo {year} {2003})}\BibitemShut
  {NoStop}%
\bibitem [{\citenamefont {Johansson}\ \emph {et~al.}(2007)\citenamefont
  {Johansson}, \citenamefont {Spar\'{e}n}, \citenamefont {Svensson},
  \citenamefont {Folestad},\ and\ \citenamefont {Claybourn}}]{Johansson2007}%
  \BibitemOpen
  \bibfield  {author} {\bibinfo {author} {\bibfnamefont {J.}~\bibnamefont
  {Johansson}}, \bibinfo {author} {\bibfnamefont {A.}~\bibnamefont
  {Spar\'{e}n}}, \bibinfo {author} {\bibfnamefont {O.}~\bibnamefont
  {Svensson}}, \bibinfo {author} {\bibfnamefont {S.}~\bibnamefont {Folestad}},
  \ and\ \bibinfo {author} {\bibfnamefont {M.}~\bibnamefont {Claybourn}},\
  }\href {\doibase 10.1366/000370207782597085} {\bibfield  {journal} {\bibinfo
  {journal} {Applied Spectroscopy}\ }\textbf {\bibinfo {volume} {61}},\
  \bibinfo {pages} {1211} (\bibinfo {year} {2007})}\BibitemShut {NoStop}%
\bibitem [{\citenamefont {Strachan}\ \emph {et~al.}(2007)\citenamefont
  {Strachan}, \citenamefont {Rades}, \citenamefont {Gordon},\ and\
  \citenamefont {Rantanen}}]{Strachan2007}%
  \BibitemOpen
  \bibfield  {author} {\bibinfo {author} {\bibfnamefont {C.~J.}\ \bibnamefont
  {Strachan}}, \bibinfo {author} {\bibfnamefont {T.}~\bibnamefont {Rades}},
  \bibinfo {author} {\bibfnamefont {K.~C.}\ \bibnamefont {Gordon}}, \ and\
  \bibinfo {author} {\bibfnamefont {J.}~\bibnamefont {Rantanen}},\ }\href
  {\doibase 10.1211/jpp.59.2.0005} {\bibfield  {journal} {\bibinfo  {journal}
  {Journal of Pharmacy and Pharmacology}\ }\textbf {\bibinfo {volume} {59}},\
  \bibinfo {pages} {179} (\bibinfo {year} {2007})}\BibitemShut {NoStop}%
\bibitem [{\citenamefont {Bell}\ and\ \citenamefont
  {Sirimuthu}(2008)}]{Bell2008}%
  \BibitemOpen
  \bibfield  {author} {\bibinfo {author} {\bibfnamefont {S.~E.~J.}\
  \bibnamefont {Bell}}\ and\ \bibinfo {author} {\bibfnamefont {N.~M.~S.}\
  \bibnamefont {Sirimuthu}},\ }\href {\doibase 10.1039/b705965p} {\bibfield
  {journal} {\bibinfo  {journal} {Chemical Society Reviews}\ }\textbf {\bibinfo
  {volume} {37}},\ \bibinfo {pages} {1012} (\bibinfo {year}
  {2008})}\BibitemShut {NoStop}%
\bibitem [{\citenamefont {Duraipandian}\ \emph {et~al.}(2018)\citenamefont
  {Duraipandian}, \citenamefont {Knopp}, \citenamefont {Pollard}, \citenamefont
  {Kerdoncuff}, \citenamefont {Petersen},\ and\ \citenamefont
  {M\"{u}llertz}}]{Duraipandian2018}%
  \BibitemOpen
  \bibfield  {author} {\bibinfo {author} {\bibfnamefont {S.}~\bibnamefont
  {Duraipandian}}, \bibinfo {author} {\bibfnamefont {M.~M.}\ \bibnamefont
  {Knopp}}, \bibinfo {author} {\bibfnamefont {M.~R.}\ \bibnamefont {Pollard}},
  \bibinfo {author} {\bibfnamefont {H.}~\bibnamefont {Kerdoncuff}}, \bibinfo
  {author} {\bibfnamefont {J.~C.}\ \bibnamefont {Petersen}}, \ and\ \bibinfo
  {author} {\bibfnamefont {A.}~\bibnamefont {M\"{u}llertz}},\ }\href {\doibase
  10.1039/C8AY00753E} {\bibfield  {journal} {\bibinfo  {journal} {Analytical
  Methods}\ }\textbf {\bibinfo {volume} {10}},\ \bibinfo {pages} {3589}
  (\bibinfo {year} {2018})}\BibitemShut {NoStop}%
\bibitem [{AST(2014)}]{ASTM}%
  \BibitemOpen
  \href@noop {} {\enquote {\bibinfo {title} {{ASTM E1840-96(2014), Standard
  Guide for Raman Shift Standards for Spectrometer Calibration}},}\ }\bibinfo
  {howpublished} {ASTM International, West Conshohocken, PA} (\bibinfo {year}
  {2014})\BibitemShut {NoStop}%
\bibitem [{\citenamefont {Hutsebaut}\ \emph {et~al.}(2005)\citenamefont
  {Hutsebaut}, \citenamefont {Vandenabeele},\ and\ \citenamefont
  {Moens}}]{Hutsebaut2005}%
  \BibitemOpen
  \bibfield  {author} {\bibinfo {author} {\bibfnamefont {D.}~\bibnamefont
  {Hutsebaut}}, \bibinfo {author} {\bibfnamefont {P.}~\bibnamefont
  {Vandenabeele}}, \ and\ \bibinfo {author} {\bibfnamefont {L.}~\bibnamefont
  {Moens}},\ }\href {\doibase 10.1039/b503624k} {\bibfield  {journal} {\bibinfo
   {journal} {The Analyst}\ }\textbf {\bibinfo {volume} {130}},\ \bibinfo
  {pages} {1024} (\bibinfo {year} {2005})}\BibitemShut {NoStop}%
\bibitem [{\citenamefont {Rodriguez}\ \emph {et~al.}(2011)\citenamefont
  {Rodriguez}, \citenamefont {Westenberger}, \citenamefont {Buhse},\ and\
  \citenamefont {Kauffman}}]{Rodriguez2011}%
  \BibitemOpen
  \bibfield  {author} {\bibinfo {author} {\bibfnamefont {J.~D.}\ \bibnamefont
  {Rodriguez}}, \bibinfo {author} {\bibfnamefont {B.~J.}\ \bibnamefont
  {Westenberger}}, \bibinfo {author} {\bibfnamefont {L.~F.}\ \bibnamefont
  {Buhse}}, \ and\ \bibinfo {author} {\bibfnamefont {J.~F.}\ \bibnamefont
  {Kauffman}},\ }\href {\doibase 10.1039/c1an15636e} {\bibfield  {journal}
  {\bibinfo  {journal} {The Analyst}\ }\textbf {\bibinfo {volume} {136}},\
  \bibinfo {pages} {4232} (\bibinfo {year} {2011})}\BibitemShut {NoStop}%
\bibitem [{\citenamefont {Bocklitz}\ \emph {et~al.}(2015)\citenamefont
  {Bocklitz}, \citenamefont {Dörfer}, \citenamefont {Heinke}, \citenamefont
  {Schmitt},\ and\ \citenamefont {Popp}}]{Bocklitz2015}%
  \BibitemOpen
  \bibfield  {author} {\bibinfo {author} {\bibfnamefont {T.~W.}\ \bibnamefont
  {Bocklitz}}, \bibinfo {author} {\bibfnamefont {T.}~\bibnamefont {Dörfer}},
  \bibinfo {author} {\bibfnamefont {R.}~\bibnamefont {Heinke}}, \bibinfo
  {author} {\bibfnamefont {M.}~\bibnamefont {Schmitt}}, \ and\ \bibinfo
  {author} {\bibfnamefont {J.}~\bibnamefont {Popp}},\ }\href {\doibase
  10.1016/j.saa.2015.04.079} {\bibfield  {journal} {\bibinfo  {journal}
  {Spectrochimica Acta Part A: Molecular and Biomolecular Spectroscopy}\
  }\textbf {\bibinfo {volume} {149}},\ \bibinfo {pages} {544} (\bibinfo {year}
  {2015})}\BibitemShut {NoStop}%
\bibitem [{\citenamefont {Liu}\ \emph {et~al.}(2018)\citenamefont {Liu},
  \citenamefont {Byrne}, \citenamefont {O'Neill},\ and\ \citenamefont
  {Hennelly}}]{Liu2018}%
  \BibitemOpen
  \bibfield  {author} {\bibinfo {author} {\bibfnamefont {D.}~\bibnamefont
  {Liu}}, \bibinfo {author} {\bibfnamefont {H.~J.}\ \bibnamefont {Byrne}},
  \bibinfo {author} {\bibfnamefont {L.}~\bibnamefont {O'Neill}}, \ and\
  \bibinfo {author} {\bibfnamefont {B.}~\bibnamefont {Hennelly}},\ }\href
  {\doibase 10.1117/12.2307574} {\bibfield  {journal} {\bibinfo  {journal}
  {Proceedings of SPIE}\ }\textbf {\bibinfo {volume} {10680}},\ \bibinfo
  {pages} {1068027} (\bibinfo {year} {2018})}\BibitemShut {NoStop}%
\bibitem [{\citenamefont {Cheng}\ and\ \citenamefont {Xie}(2013)}]{Cheng2013}%
  \BibitemOpen
  \bibinfo {editor} {\bibfnamefont {J.-X.}\ \bibnamefont {Cheng}}\ and\
  \bibinfo {editor} {\bibfnamefont {X.~S.}\ \bibnamefont {Xie}},\ eds.,\
  \href@noop {} {\emph {\bibinfo {title} {Coherent Raman Scattering
  Microscopy}}}\ (\bibinfo  {publisher} {CRC Press, Taylor \& Francis Group},\
  \bibinfo {address} {Boca Raton, USA},\ \bibinfo {year} {2013})\BibitemShut
  {NoStop}%
\bibitem [{\citenamefont {K\"{o}nig}\ \emph {et~al.}(1995)\citenamefont
  {K\"{o}nig}, \citenamefont {Liang}, \citenamefont {Berns},\ and\
  \citenamefont {Tromberg}}]{Konig1995}%
  \BibitemOpen
  \bibfield  {author} {\bibinfo {author} {\bibfnamefont {K.}~\bibnamefont
  {K\"{o}nig}}, \bibinfo {author} {\bibfnamefont {H.}~\bibnamefont {Liang}},
  \bibinfo {author} {\bibfnamefont {M.~W.}\ \bibnamefont {Berns}}, \ and\
  \bibinfo {author} {\bibfnamefont {B.~J.}\ \bibnamefont {Tromberg}},\ }\href
  {\doibase 10.1038/377020a0} {\bibfield  {journal} {\bibinfo  {journal}
  {Nature}\ }\textbf {\bibinfo {volume} {377}},\ \bibinfo {pages} {20}
  (\bibinfo {year} {1995})}\BibitemShut {NoStop}%
\bibitem [{\citenamefont {K\"{o}nig}\ \emph {et~al.}(1999)\citenamefont
  {K\"{o}nig}, \citenamefont {Becker}, \citenamefont {Fischer}, \citenamefont
  {Riemann},\ and\ \citenamefont {Halbhuber}}]{Konig1999}%
  \BibitemOpen
  \bibfield  {author} {\bibinfo {author} {\bibfnamefont {K.}~\bibnamefont
  {K\"{o}nig}}, \bibinfo {author} {\bibfnamefont {T.~W.}\ \bibnamefont
  {Becker}}, \bibinfo {author} {\bibfnamefont {P.}~\bibnamefont {Fischer}},
  \bibinfo {author} {\bibfnamefont {I.}~\bibnamefont {Riemann}}, \ and\
  \bibinfo {author} {\bibfnamefont {K.-J.}\ \bibnamefont {Halbhuber}},\ }\href
  {\doibase 10.1364/OL.24.000113} {\bibfield  {journal} {\bibinfo  {journal}
  {optics Letters}\ }\textbf {\bibinfo {volume} {24}},\ \bibinfo {pages} {113}
  (\bibinfo {year} {1999})}\BibitemShut {NoStop}%
\bibitem [{\citenamefont {Fu}\ \emph {et~al.}(2006)\citenamefont {Fu},
  \citenamefont {Wang}, \citenamefont {Shi},\ and\ \citenamefont
  {Cheng}}]{Fu2006}%
  \BibitemOpen
  \bibfield  {author} {\bibinfo {author} {\bibfnamefont {Y.}~\bibnamefont
  {Fu}}, \bibinfo {author} {\bibfnamefont {H.}~\bibnamefont {Wang}}, \bibinfo
  {author} {\bibfnamefont {R.}~\bibnamefont {Shi}}, \ and\ \bibinfo {author}
  {\bibfnamefont {J.-X.}\ \bibnamefont {Cheng}},\ }\href {\doibase
  10.1364/OE.14.003942} {\bibfield  {journal} {\bibinfo  {journal} {Optics
  Express}\ }\textbf {\bibinfo {volume} {14}},\ \bibinfo {pages} {3942}
  (\bibinfo {year} {2006})}\BibitemShut {NoStop}%
\bibitem [{\citenamefont {Owyoung}\ and\ \citenamefont
  {Jones}(1977)}]{Owyoung1977}%
  \BibitemOpen
  \bibfield  {author} {\bibinfo {author} {\bibfnamefont {A.}~\bibnamefont
  {Owyoung}}\ and\ \bibinfo {author} {\bibfnamefont {E.~D.}\ \bibnamefont
  {Jones}},\ }\href {\doibase 10.1364/OL.1.000152} {\bibfield  {journal}
  {\bibinfo  {journal} {Optics Letters}\ }\textbf {\bibinfo {volume} {1}},\
  \bibinfo {pages} {152} (\bibinfo {year} {1977})}\BibitemShut {NoStop}%
\bibitem [{\citenamefont {Owyoung}(1978)}]{Owyoung1978}%
  \BibitemOpen
  \bibfield  {author} {\bibinfo {author} {\bibfnamefont {A.}~\bibnamefont
  {Owyoung}},\ }\href {\doibase 10.1364/OL.2.000091} {\bibfield  {journal}
  {\bibinfo  {journal} {Optics Letters}\ }\textbf {\bibinfo {volume} {2}},\
  \bibinfo {pages} {91} (\bibinfo {year} {1978})}\BibitemShut {NoStop}%
\bibitem [{\citenamefont {Dom\'{e}nech}\ and\ \citenamefont
  {Cueto}(2013)}]{Domenech2013}%
  \BibitemOpen
  \bibfield  {author} {\bibinfo {author} {\bibfnamefont {J.~L.}\ \bibnamefont
  {Dom\'{e}nech}}\ and\ \bibinfo {author} {\bibfnamefont {M.}~\bibnamefont
  {Cueto}},\ }\href {\doibase 10.1364/OL.38.004074} {\bibfield  {journal}
  {\bibinfo  {journal} {Optics Letters}\ }\textbf {\bibinfo {volume} {38}},\
  \bibinfo {pages} {4074} (\bibinfo {year} {2013})}\BibitemShut {NoStop}%
\bibitem [{\citenamefont {Westergaard}\ \emph {et~al.}(2015)\citenamefont
  {Westergaard}, \citenamefont {Lassen},\ and\ \citenamefont
  {Petersen}}]{Westergaard2015}%
  \BibitemOpen
  \bibfield  {author} {\bibinfo {author} {\bibfnamefont {P.~G.}\ \bibnamefont
  {Westergaard}}, \bibinfo {author} {\bibfnamefont {M.}~\bibnamefont {Lassen}},
  \ and\ \bibinfo {author} {\bibfnamefont {J.~C.}\ \bibnamefont {Petersen}},\
  }\href {\doibase 10.1364/OE.23.016320} {\bibfield  {journal} {\bibinfo
  {journal} {Optics Express}\ }\textbf {\bibinfo {volume} {23}},\ \bibinfo
  {pages} {16320} (\bibinfo {year} {2015})}\BibitemShut {NoStop}%
\bibitem [{\citenamefont {Meng}\ \emph {et~al.}(2013)\citenamefont {Meng},
  \citenamefont {Petrov},\ and\ \citenamefont {Yakovlev}}]{Meng2013}%
  \BibitemOpen
  \bibfield  {author} {\bibinfo {author} {\bibfnamefont {Z.}~\bibnamefont
  {Meng}}, \bibinfo {author} {\bibfnamefont {G.~I.}\ \bibnamefont {Petrov}}, \
  and\ \bibinfo {author} {\bibfnamefont {V.~V.}\ \bibnamefont {Yakovlev}},\
  }\href {\doibase 10.1088/1612-2011/10/6/065701} {\bibfield  {journal}
  {\bibinfo  {journal} {Laser Physics Letters}\ }\textbf {\bibinfo {volume}
  {10}},\ \bibinfo {pages} {065701} (\bibinfo {year} {2013})}\BibitemShut
  {NoStop}%
\bibitem [{\citenamefont {Hu}\ \emph {et~al.}(2013)\citenamefont {Hu},
  \citenamefont {Slipchenko}, \citenamefont {Wang}, \citenamefont {Wang},
  \citenamefont {Lin}, \citenamefont {Simpson}, \citenamefont {Hu},\ and\
  \citenamefont {Cheng}}]{Hu2013}%
  \BibitemOpen
  \bibfield  {author} {\bibinfo {author} {\bibfnamefont {C.-R.}\ \bibnamefont
  {Hu}}, \bibinfo {author} {\bibfnamefont {M.~N.}\ \bibnamefont {Slipchenko}},
  \bibinfo {author} {\bibfnamefont {P.}~\bibnamefont {Wang}}, \bibinfo {author}
  {\bibfnamefont {P.}~\bibnamefont {Wang}}, \bibinfo {author} {\bibfnamefont
  {J.~D.}\ \bibnamefont {Lin}}, \bibinfo {author} {\bibfnamefont
  {G.}~\bibnamefont {Simpson}}, \bibinfo {author} {\bibfnamefont
  {B.}~\bibnamefont {Hu}}, \ and\ \bibinfo {author} {\bibfnamefont {J.-X.}\
  \bibnamefont {Cheng}},\ }\href {\doibase 10.1364/OL.38.001479} {\bibfield
  {journal} {\bibinfo  {journal} {Optics Letters}\ }\textbf {\bibinfo {volume}
  {38}},\ \bibinfo {pages} {1479} (\bibinfo {year} {2013})}\BibitemShut
  {NoStop}%
\bibitem [{\citenamefont {Barrett}\ and\ \citenamefont
  {Begley}(1975)}]{Barrett1975}%
  \BibitemOpen
  \bibfield  {author} {\bibinfo {author} {\bibfnamefont {J.~J.}\ \bibnamefont
  {Barrett}}\ and\ \bibinfo {author} {\bibfnamefont {R.~F.}\ \bibnamefont
  {Begley}},\ }\href {\doibase 10.1063/1.88379} {\bibfield  {journal} {\bibinfo
   {journal} {Applied Physics Letters}\ }\textbf {\bibinfo {volume} {27}},\
  \bibinfo {pages} {129} (\bibinfo {year} {1975})}\BibitemShut {NoStop}%
\bibitem [{\citenamefont {Akhmanov}\ \emph {et~al.}(1978)\citenamefont
  {Akhmanov}, \citenamefont {Gadzhiev}, \citenamefont {Koroteev}, \citenamefont
  {Orlov},\ and\ \citenamefont {Shumay}}]{Akhmanov1978}%
  \BibitemOpen
  \bibfield  {author} {\bibinfo {author} {\bibfnamefont {S.~A.}\ \bibnamefont
  {Akhmanov}}, \bibinfo {author} {\bibfnamefont {F.~N.}\ \bibnamefont
  {Gadzhiev}}, \bibinfo {author} {\bibfnamefont {N.~I.}\ \bibnamefont
  {Koroteev}}, \bibinfo {author} {\bibfnamefont {R.~Y.}\ \bibnamefont {Orlov}},
  \ and\ \bibinfo {author} {\bibfnamefont {I.~L.}\ \bibnamefont {Shumay}},\
  }\href@noop {} {\bibfield  {journal} {\bibinfo  {journal} {Soviet Journal of
  Experimental and Theoretical Physics Letters}\ }\textbf {\bibinfo {volume}
  {27}},\ \bibinfo {pages} {243} (\bibinfo {year} {1978})}\BibitemShut
  {NoStop}%
\bibitem [{\citenamefont {Nielsen}(2002)}]{Nielsen2002}%
  \BibitemOpen
  \bibfield  {author} {\bibinfo {author} {\bibfnamefont {L.}~\bibnamefont
  {Nielsen}},\ }in\ \href@noop {} {\emph {\bibinfo {booktitle} {Algorithms for
  Approximation IV}}},\ \bibinfo {editor} {edited by\ \bibinfo {editor}
  {\bibfnamefont {J.}~\bibnamefont {Levesley}}, \bibinfo {editor}
  {\bibfnamefont {I.}~\bibnamefont {Anderson}}, \ and\ \bibinfo {editor}
  {\bibfnamefont {J.~C.}\ \bibnamefont {Mason}}}\ (\bibinfo  {publisher}
  {University of Huddersfield, Huddersfield, UK},\ \bibinfo {year}
  {2002})\BibitemShut {NoStop}%
\bibitem [{\citenamefont {Ciddor}(1996)}]{Ciddor1996}%
  \BibitemOpen
  \bibfield  {author} {\bibinfo {author} {\bibfnamefont {P.~E.}\ \bibnamefont
  {Ciddor}},\ }\href {\doibase 10.1364/ao.35.001566} {\bibfield  {journal}
  {\bibinfo  {journal} {Applied Optics}\ }\textbf {\bibinfo {volume} {35}},\
  \bibinfo {pages} {1566} (\bibinfo {year} {1996})}\BibitemShut {NoStop}%
\bibitem [{\citenamefont {BIPM}\ \emph {et~al.}(2008)\citenamefont {BIPM},
  \citenamefont {IEC}, \citenamefont {IFCC}, \citenamefont {ILAC},
  \citenamefont {ISO}, \citenamefont {IUPAC}, \citenamefont {IUPAP},\ and\
  \citenamefont {OIML}}]{GUM}%
  \BibitemOpen
  \bibfield  {author} {\bibinfo {author} {\bibnamefont {BIPM}}, \bibinfo
  {author} {\bibnamefont {IEC}}, \bibinfo {author} {\bibnamefont {IFCC}},
  \bibinfo {author} {\bibnamefont {ILAC}}, \bibinfo {author} {\bibnamefont
  {ISO}}, \bibinfo {author} {\bibnamefont {IUPAC}}, \bibinfo {author}
  {\bibnamefont {IUPAP}}, \ and\ \bibinfo {author} {\bibnamefont {OIML}},\
  }\href@noop {} {\enquote {\bibinfo {title} {{Guide to the Expression of
  Uncertainty in Measurement}},}\ }\bibinfo {howpublished} {JCGM 100:2008, GUM
  1995 with minor corrections} (\bibinfo {year} {2008})\BibitemShut {NoStop}%
\end{thebibliography}%

\end{document}